\documentclass{iaus}
\usepackage{graphicx}
\def\mc{meridional circulation}
\def\pf{poloidal field}
\def\mm{Maunder minimum}
\def\gm{grand minima}
\def\fl{fluctuations}
\def\bl{Babcock-Leighton}
\def\ftdm{flux transport dynamo model}

\title[Modelling grand minima using a dynamo model] 
{Modelling grand minima of solar activity using a flux transport dynamo model}
\author[Bidya Binay Karak \& Arnab Rai Choudhuri]   
{Bidya Binay Karak$^1$
\and Arnab Rai Choudhuri}
\affiliation{Department of Physics, Indian Institute of Science, Bangalore 560012, India\\$^1$email: {\tt bidya\_karak@physics.iisc.ernet.in}}
\pubyear{2013}
\volume{294}  
\jname{Solar and Astrophysical Dynamos and Magnetic Activity}
\editors{A.G. Kosovichev, E.M. de Gouveia Dal Pino, \& Y.Yan, eds.}
\begin{document}
\maketitle
\begin{abstract}
The occurrence of grand minima like the Maunder minimum is an intriguing aspect
of the sunspot cycle.  We use the flux transport dynamo model to explain the grand
minima, showing that they arise when either the poloidal field or the \mc\ falls to a sufficiently
low value due to fluctuations.  Assuming these fluctuations to be Gaussian and
determining the various parameters from the data of the last 28 cycles, we carry
on a dynamo simulation with both these fluctuations.  The results are remarkably close
to the observational data.
\end{abstract}
\firstsection 
\section{Introduction}
Early observations of sunspots revealed that there was an epoch from 1645 to 1715 when 
sunspots almost disappeared from the surface of the Sun. This period is known as 
the \mm. This was not an artifact of very few observations as this period was well covered 
by direct observations (Hoyt \& Schatten 1996). Several old archival data have shown that there 
was a strong north-south asymmetry during the last phase of the \mm\ because most of the 
sunspots appeared in the southern hemisphere of the Sun (Ribes \& Nesme-Ribes 1993).
From the study of cosmogenic isotopes $^{10}$Be and $^{14}$C, it was found that the solar activity
was weaker during the \mm, but the cyclic oscillation of solar activity continued with periods 
longer than the usual 11 years (Miyahara {\it et al.} 2004).
The production of these isotopes varies with the solar cycle because the strong magnetic 
field of the Sun during solar maximum suppresses the Earth-ward incoming cosmic rays flux.
The study of cosmogenic isotopes revealed that the \mm, which has been seen in the direct
observational data during the seventeenth century, was not unique (Usoskin, Solanki \& Kovaltsov 2007; Nagaya et al. 2012). 
In particular, from the study of $^{14}$C data, Usoskin, Solanki \& Kovaltsov (2007) 
reported that there have been about 27 grand minima in the last 11,000 years and 
the Sun spent about 17\% of this time in the grand minima state.

Our motivation is to study the physics behind the origin of these grand minima and then to find out
their frequency using a dynamo model. Our calculations are based on the
flux transport dynamo model, which is the most promising model at present to study the
solar cycle and of which a version has been developed in our group over the years 
(Choudhuri, Sch\"ussler \& Dikpati 1995; Nandy \& Choudhuri 2002; Chatterjee, Nandy \& Choudhuri 2004; 
Chatterjee \& Choudhuri 2006; Goel \& Choudhuri 2009; 
Karak \& Choudhuri 2012; Karak \& Petrovay 2013; Karak \& Nandy 2012). 
The details of this model have been reviewed by Choudhuri (2011). Basically
this is a kinematic mean-field dynamo model in which the strong toroidal field is generated near the
base of the convection zone by the differential rotation and the poloidal field is generated 
from the decay of the tilted bipolar sunspots near the solar surface (\bl\ process). The \mc\
and the turbulent diffusivity provide the two important transport processes in this model
for transporting the poloidal field from the top of the convection zone to the bottom.

\begin{figure}[!h]
\begin{center}
\includegraphics[width=4.0in]{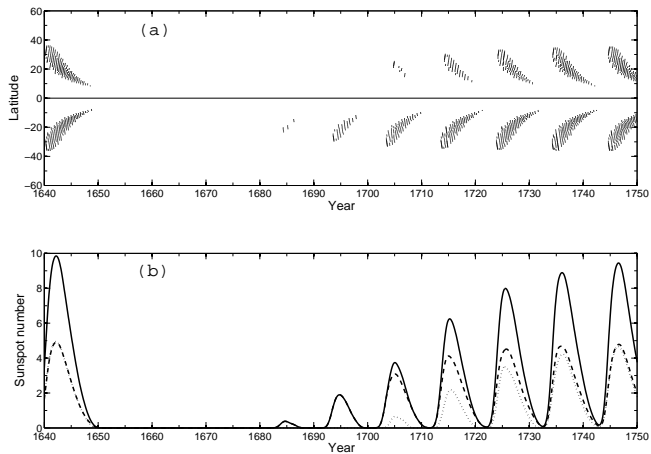} 
\caption{Theoretical Maunder minimum produced by making the poloidal field weak. (a) The butterfly
diagram. (b) The smoothed sunspot number. The dashed and dotted lines show the
sunspot numbers in southern and northern hemispheres, whereas the solid line is the 
total sunspot number. From Choudhuri \& Karak (2009).}
\label{mm_pol}
\end{center}
\end{figure}

\section{Modelling the Maunder minimum}
Let us begin by looking at the possible 
sources of \fl\ in the \ftdm\ which can produce grand minima. A detailed discussion of our
present understanding of the mechanisms that may cause grand minima has been reviewed by Choudhuri (2012).
Therefore we are not going into the same discussion again. We believe that the possible 
mechanisms for producing the grand minima are the following: 
{\it i}) \fl\ in the \bl\ (B-L) process that may make the \pf\ weak and
{\it ii}) \fl\ in the \mc\ that may make it slow. 

Let us look at the B-L process first. 
In this process, the poloidal field is generated from the decay of the 
tilted bipolar sunspots. This is a complex nonlinear process which depends 
on many factors, such as tilt angles of the bipolar sunspots. It has been found 
that the average tilt angle varies cycle to cycle making the \pf\
different at every solar minimum from its mean value (Dasi-Espuig {\it et al.} 2010). 
While the tilts of bipolar sunspots are produced by the Coriolis force
acting on the rising toroidal flux tubes (D'Silva \& Choudhuri 1993), the fluctuations in the tilts 
presumably result from the buffeting of the flux tubes by turbulence (Longcope \& Choudhuri 2002).
These fluctuations in tilts introduce fluctuations in the B-L process that
has been identified as a main source of irregularities in solar cycles (Choudhuri, 
Chatterjee \& Jiang 2007; Jiang, Chatterjee \& Choudhuri 2007).
Several authors (Choudhuri 1992; Charbonneau, Blais-Laurier \& St-Jean 2004, Passos \& Lopes 2011) 
introduced fluctuations in the \pf\ 
generation process and found intermittencies resembling grand minima.
We model the \mm\ with the assumption that the poloidal field before the \mm\ dropped to a very low
value. All our studies are based on the \ftdm\ presented by Chatterjee, Nandy \& Choudhuri (2004), with 
a few parameters modified by Karak (2010). To produce the \mm, we (Choudhuri \& Karak 2009) did the following.
We stop the code at a solar minimum and change the poloidal field by a factor $\gamma$. 
We take $\gamma = 0.4$ in southern hemisphere and $0.0$ in northern hemisphere.
After this modification in the \pf, we again run the model for several solar cycles. 
Figure \ref{mm_pol} shows the result of this procedure. Note that for the sake of 
comparison with the observation data, we have marked the beginning of this figure 
to be the year 1640. Our results successfully reproduce the north-south asymmetry 
in the sunspots and the sudden initiation of \mm\ and the gradual recovery.

\begin{figure}
\begin{center}
 \includegraphics[width=4.0in]{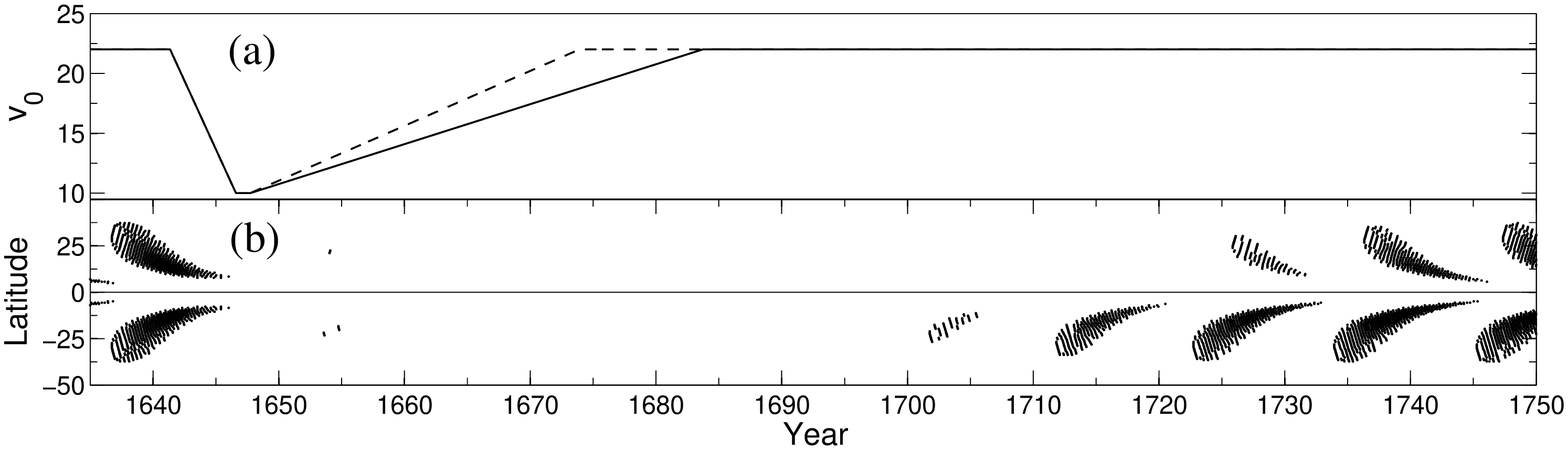}
\vspace{-4.0cm}
\caption{(a) The solid and dashed lines show the variations of $v_0$ (in m~s$^{-1}$) in northern and southern 
hemispheres with time. (b) Butterfly diagram during the Maunder minimum. From Karak (2010).}
   \label{mm_mc}
\end{center}
\end{figure}

The other important source of \fl\ in this dynamo model is the variations of the
meridional circulation. There is no long-term observational data of the \mc\ in the past.
Therefore we do not directly know whether the \mc\ had large variations with time. However,
we know that there have been variations in the periods of solar cycles.
Since the period of the flux transport dynamo strongly depends on the strength of
the meridional circulation (Yeates, Nandy \& Mackey 2008), we
believe that the variations in the periods of the past cycles imply variations
in the \mc. Karak \& Choudhuri (2011) used this idea to get a rough 
idea about the variations of \mc. They also showed that such 
temporal variations in the \mc\ are necessary to model the {\it Waldmeier effect}.
To produce the \mm, Karak (2010) decreased the amplitude of \mc\ rapidly to a very low value during a cycle and 
then, after a few years, it was increased to the usual value. By doing this, he is able to 
reproduce the \mm\ remarkably well. Figure \ref{mm_mc} shows this result.
The physics of this result can be understood from Yeates, Nandy \& Mackey (2008) 
in the following way. When we decrease the \mc\ to a very low value, the solar cycle periods become longer
and the \pf\ remains in the convection zone for longer time. 
There are two competing effects -- {\it i}) the differential rotation gets more time to 
induct the toroidal field which makes the toroidal field stronger and {\it ii}) diffusion 
gets more time to diffuse the fields which makes the toroidal field weaker. Now, in our 
high diffusivity model, the latter effect (the diffusion of the fields) is more important. 
A sufficiently weak \mc\ can make the toroidal field so weak that the dynamo
is pushed into a grand minimum state.

\section{Estimating the probability of grand minima}
\begin{figure}
\begin{center}
 \includegraphics[width=2.5in]{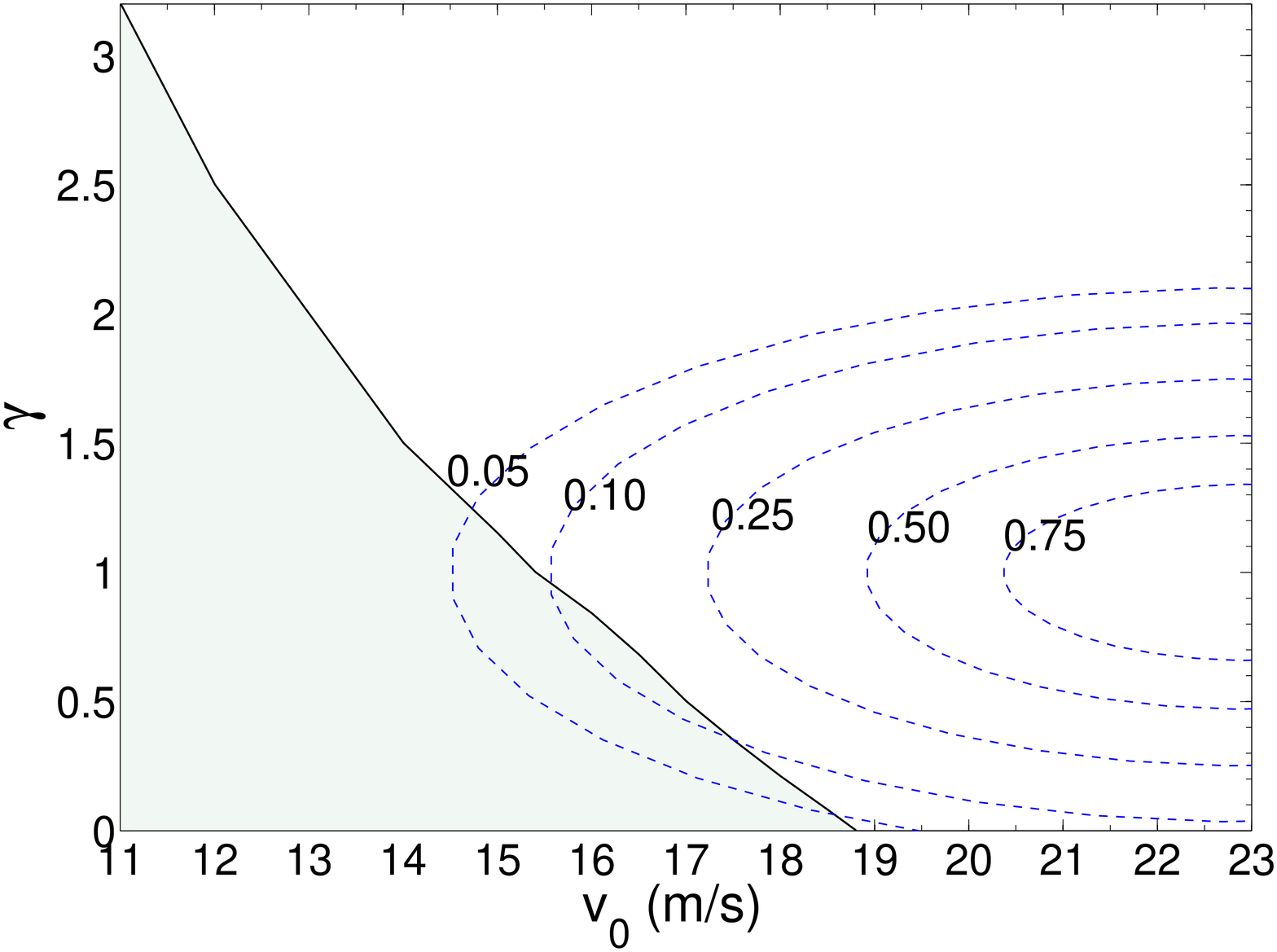}
 \caption{The solid line shows the values of the \mc\ amplitude $v_0$ and the poloidal
field scale factor $\gamma$ which produce grand minima of duration
$\sim 20$~yrs. The parameters lying in the shaded region produce grand minima of
longer duration. The dashed curves are the contours of the joint probability
$P(\gamma, v_0)$, with the values of $P(\gamma, v_0)$ (excluding the constant pre-factor) given in the plot. From Choudhuri \& Karak (2012).}
   \label{parameters}
\end{center}
\end{figure}

\begin{figure}
\begin{center}
 \includegraphics[width=3.0in]{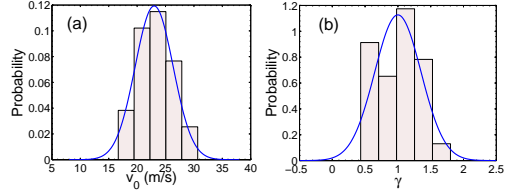}
 \caption{Histograms of the derived strength of (a) the \mc\ $v_0$ and (b) the \pf\.
The solid curves are the Gaussian functions with means and standard deviations 23 m~s$^{-1}$ 3.34 m~s$^{-1}$ 
in (a) and 1 and $0.35$ in (b). From Choudhuri \& Karak (2012).}
\label{hists}
\end{center}
\end{figure}

In the observation data of last 11,000~yrs, there were 27 grand minima with 
durations longer than about 20~yrs. Therefore the probability that a particular solar cycle triggers
a grand minimum is $2.7\%$. Our present motivation is to see whether the existing \ftdm\
can give a correct value of this probability. In the previous section, we have seen that the \mm\
can be reproduced either by decreasing the poloidal field to a very low value
or by decreasing the meridional circulation 
to a very low value. However, instead of reducing either of these two quantities to a very low value,
if we reduce both simultaneously, then we can reproduce the \mm\ for more moderate values of these two 
quantities. Karak (2010) gave the required values of the strength of the \mc\ $v_0$ 
and the strength of the \pf\ $\gamma$ which can produce a Maunder-like grand minimum
(see Fig.~6 of Karak 2010). Here we present a similar plot in Fig.\ \ref{parameters}.
Parameters lying on the solid line produce grand minima of duration 
about 20~yrs. If the parameters lie below the solid line, then we get grand
minima of durations more than 20~yrs. The shaded region in 
Fig.\ \ref{parameters} should, therefore, cover all the grand minima observed in the past. 
Now the question is how to find out the values of these $v_0$ and $\gamma$ in the past. 
Obviously there is no direct way to find out these values. We (Choudhuri \& Karak 2012) got these values 
in indirect ways. We find the values of $v_0$ by 
assuming that $v_0$ goes inversely as the solar cycle period. 
Therefore, from the observed periods of the last 28 solar cycles, we 
get 28 data for the $v_0$ and  construct the histogram shown in Fig.\ \ref{hists}(a).
The Gaussian fit to this 
histogram is shown by the solid line in Fig.\ \ref{hists}(a). Next we find out 
the values of the \pf\ by assuming a perfect correlation between the peak sunspot number 
and the \pf\ strength of the previous cycle. Therefore we again get 28
data points for $\gamma$ from the peak sunspot numbers of last 28 solar cycles. 
The histogram is shown in Fig.\ \ref{hists}(b), with a Gaussian fit. 
The joint probability density that $v_0$ and 
$\gamma$ of a cycle lie in the range $v_0$, $v_0 + $d$v_0$ 
and $\gamma$, $\gamma + $d$\gamma$ is given by
$$P (\gamma, v_0) d\gamma dv_0 = \frac{1}{\sigma_v \sqrt{2 \pi}} \exp\left[- \frac{(v_0 - \overline{v_0})^2}{2 \sigma_v^2}\right] \times \frac{1}{\sigma_{\gamma} \sqrt{2 \pi}} \exp\left[- \frac{(\gamma - 1)^2}{2 \sigma_{\gamma}^2}\right] d\gamma \, dv_0.$$ 
By integrating the above probability density over the shaded region in Fig.\ \ref{parameters}, 
we get the probability of a solar cycle triggering a grand minimum as $1.3\%$, 
which is not very far from the observed value $2.7\%$.

\section{Simulations of grand minima}
\begin{figure}[!h]
\begin{center}
 \includegraphics[width=5.0in]{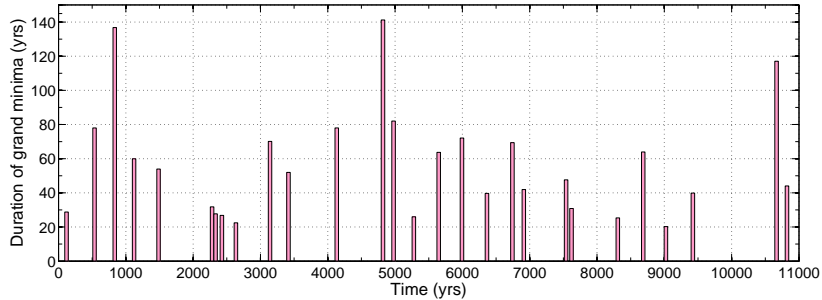}
 \caption{The durations of grand minima indicated by vertical bars at their times of occurrence in a 11,000~yr simulation.
This is the result of a particular realization of random fluctuations that produced 28 grand minima. From Choudhuri \& Karak (2012).}
   \label{grand_min}
\end{center}
\end{figure}

To check whether the above simple argument about the occurrence probability of grand minima
is borne out by detailed calculations, we (Choudhuri \& Karak 2012; Karak \& Choudhuri 2013) have carried out 
extensive simulations with our dynamo model.
We introduce stochastic fluctuations both in the poloidal field and the \mc.
To introduce fluctuations in the poloidal field, we change the \pf\ by the factor $\gamma$ 
above $0.8R_{\odot}$ 
at every solar minima (for details see Choudhuri, Chatterjee \& Jiang 2007),
whereas to introduce fluctuations in the \mc, we change $v_0$ everywhere in the model 
after a certain coherence time of 30~yrs. We choose the fluctuation levels of the \pf\ 
and the \mc\ according to their distributions shown in Fig.\ \ref{hists}.
A result of a typical run of 11,000~yrs is shown in Fig.\ \ref{grand_min}. In this particular run,
we get 28 grand minima. We have carried out several simulations with different realizations of the
random numbers for the \pf\ and the \mc, finding that the numbers of grand minima in all simulations lie in the
range 24-30 remarkably close to the observed number 27.
Another important result we find is that the Sun spent about 10--15\% of the total time 
in grand minima state which is close to the observed value 17\%.
We have also checked how the results change with the coherence time of \mc\ by performing 
several simulations with varying coherence time. We have seen that when we take the 
coherence time less than about 15~yrs, the number of \gm\ becomes considerably less.

One important point to note is that the B-L process may not work during \gm,
since very few sunspots appear during those times. Therefore the question remains 
unclear how the Sun recovers from a grand minimum after entering one. 
We believe that the turbulent $\alpha$ effect 
originally proposed by Parker (1955) might be a good candidate for generating a 
weak poloidal field and this could eventually pull the Sun out of the grand minimum. 
Since the detailed nature of this $\alpha$ effect is unclear, our first calculations
described here use the same B-L $\alpha$ all the time. However our future work 
(Karak \& Choudhuri 2013) explores this issue.

\section{Conclusion}
We have explored the possible causes of the grand minima using a \ftdm. We believe
that the fluctuations in the \pf\ and in the \mc\ are the two important sources of randomness 
in this dynamo model. We have shown that a weak poloidal field or a weak \mc\ or 
both can push the Sun into a grand minimum. We have also studied the 
frequency of occurrence of the grand minima. We have derived the levels of the 
fluctuations of the \pf\ and the \mc\ from the data of last 28 solar cycles. Then we use these 
observationally derived fluctuations in our dynamo model to simulate grand minima in a run of 11,000~yrs. 
Our result of the frequency of the grand minima is exceptionally close 
to the observational result.
The details of this work, including the recovery mechanism of grand minima, and the grand maxima 
can be found in recent work (Karak \& Choudhuri 2013).

\begin{acknowledgements}
The authors thank  Department of Science and Technology, Government of
India, for the travel support to participate this symposium
(ARC through JC Bose Fellowship).

\end{acknowledgements}


\begin{thebibliography}{}
\bibitem[Charbonneau, Blais-Laurier \& St-Jean 2004]{charbon04}
{Charbonneau, P., Blais-Laurier, G., \& St-Jean, C.} 2004, 
\textit{ApJ}, 616, L183

\bibitem[Chatterjee \& Choudhuri 2006]{cc06}
{Chatterjee, P., \& Choudhuri, A.\ R.} 2006,
{\it Solar Phys.}, 239, 29

\bibitem[Chatterjee, Nandy \& Choudhuri (2004)]{chatterjee}
{Chatterjee, P., Nandy, D. \& Choudhuri, A. R.} 2004,
\textit{A\&A}, 427, 1019

\bibitem[Choudhuri (1992)]{choudhuri92} 
{Choudhuri, A.\ R.} 1992, 
\textit{A\&A}, 253, 277

\bibitem[Choudhuri 2011]{ch11}
  Choudhuri, A.\ R. 2011, {\it Pramana}, 77, 77

\bibitem[]{cho12}
Choudhuri, A.\ R. 2012, in: C. H. Mandrini
\& D. F. Webb (eds.), \textit{Comparative Magnetic Minima: 
Characterizing quiet times in the Sun and Stars},
Proc. IAU Symposium No. 286, p.\ 350

\bibitem[Choudhuri, Chatterjee \& Jiang (2007)]{ccj}
{Choudhuri, A.\ R., Chatterjee, P., \& Jiang, J.} 2007,
\textit{Phys. Rev. Lett.}, 98, 131103

\bibitem[Choudhuri \& Karak (2009)]{karak}
{Choudhuri, A. R., \& Karak, B. B.} 2009,
\textit{RAA}, 9, 953

\bibitem[Choudhuri \& Karak (2012)]{choudhurikarak}
{Choudhuri, A.\ R. \& Karak, B.\ B.} 2012,
\textit{Phys. Rev. Lett.}, 109, 171103

\bibitem[Choudhuri, Sch\"ussler \& Dikpati (1995)]{chou95}
{Choudhuri, A. R., Sch\"ussler, M., \& Dikpati, M.} 1995,
\textit{A\&A}, 303, L29

\bibitem[D'Silva \& Choudhuri (1993)]{dsilva93} 
{D'Silva, S., \& Choudhuri, A.\ R.} 1993, 
\textit{A\&A}, 272, 621

\bibitem[Dasi-Espuig {\it et al.} (2010)]{dasi10} 
{Dasi-Espuig, M., Solanki, S.\ K., Krivova, N.\ A., Cameron, R. \& Pe\~nuela, T.} 2010,
\textit{A\&A}, 518, 7

\bibitem[Jiang, Chatterjee \& Choudhuri (2007)]{jiang}
{Jiang, J., Chatterjee, P., \& Choudhuri, A. R.} 2007,
\textit{MNRAS}, 381, 1527

\bibitem[Goel \& Choudhuri (2009)]{goel}
{Goel, A. \& Choudhuri, A.\ R.} 2009, 
\textit{\it RAA}, 9, 115

\bibitem[Hoyt \& Schatten (1996)]{hoyt}
{Hoyt, D.\ V., \& Schatten, K.\ H.} 1996, 
\textit{Solar Phys.}, 165, 181

\bibitem[Karak (2010)]{karak10}
{Karak, B. B.} 2010,
\textit{ApJ}, 724, 1021

\bibitem[Karak \& Choudhuri (2011)]{karaknew}
{Karak, B.\ B., \& Choudhuri, A.\ R.} 2011,
\textit{MNRAS}, 410, 1503

\bibitem[Karak \& Choudhuri (2012)]{karak12}
{Karak, B.\ B., \& Choudhuri, A.\ R.} 2012,
\textit{Solar Phys.}, 278, 137

\bibitem[Karak \& Choudhuri (2013)]{karak13}
{Karak, B.\ B., \& Choudhuri, A.\ R.} 2013,
\textit{\it RAA}, arXiv:1306.5438

\bibitem[Karak \& Petrovay (2013)]{karakpetrovay}
{Karak, B.\ B., \& Petrovay, K.} 2013,
\textit{Solar Phys.}, 282, 321


\bibitem[Karak \& Nandy (2012)]{karaknandy}
{Karak, B.\ B., \& Nandy, D.} 2012, 
{\it ApJ Lett.}, 761, L13

\bibitem[Longcope \& Choudhuri 2002]{lc02}
 {Longcope, D. \& Choudhuri, A.\ R} 2002
 \textit{Solar Phys.} 205, 63

\bibitem[]{key-175}
Miyahara, H., {et al.} 2004, {\it Solar Phys.}, 224, 317

\bibitem[Nagaya {\it et al.} 2012]{nagaya}
{Nagaya, K. {\it et al.}} 2012,
\textit{Solar Phys.}, 280, 223.

\bibitem[]{nan02}
Nandy, D., \& Choudhuri, A.\ R. 2002, {\it Science}, 296, 1671

\bibitem[Parker (1955)]{parker55} 
{Parker, E.\ N.} 1955, 
\textit{ApJ}, 122, 293

\bibitem[Passos \& Lopes (2011)]{pl11}Passos, D. \& Lopes, I. 2011, \textit{JASTP}, 73, 191

\bibitem[Ribes \&  Nesme-Ribes 1993]{ribes}
{Ribes, J.\ C., \& Nesme-Ribes, E.} 1993, 
\textit{A\&A}, 276, 549

\bibitem[Usoskin, Solanki \& Kovaltsov (2007)]{usos07}
 {Usoskin, I.\ G., Solanki, S.\ K., \& Kovaltsov, G.\ A.} 2007, 
 \textit{A\&A}, 471, 301

\bibitem[Yeates, Nandy \& Mackay (2008)]{yeates} 
 {Yeates, A.\ R., Nandy, D., \& Mackay, D.\ H.} 2008, 
 \textit{ApJ}, 673, 544

\end{thebibliography}
\end{document}